\documentclass{jpsj2}
\usepackage{graphicx}

\newcommand{\etal}{{\em et al}.}

\newcommand{\alf}{$\alpha$}

\newcommand{\gam}{$\gamma$}

\newcommand{\moo}{$\mu$}

\newcommand{\kB}{k$_B$}

\newcommand{\Nf}{N$_f$}

\newcommand{\nf}{n$_f$}

\newcommand{\TK}{T$_K$}

\newcommand{\epsf}{$\epsilon_f$}

\newcommand{\Uff}{U$_{ff}$}

\newcommand{\DE}{$\Delta$($\varepsilon$)}

\newcommand{\DLS}{$\Delta_{LS}$}

\newcommand{\VE}{V($\varepsilon$)}

\newcommand{\hnu}{$h\nu$}

\newcommand{\spec}{$\rho_{LOC}(\omega)$}

\newcommand{\specf}{$\rho_f(\omega)$}

\newcommand{\specfLOC}{$\rho_{fLOC}(\omega)$}

\newcommand{\specfNON}{$\rho_{fNON}(\omega)$}

\newcommand{\specfNONEF}{$\rho_{fNON}$(\EF)}

\newcommand{\specfLOCEF}{$\rho_{fLOC}$(\EF)}

\newcommand{\specfEF}{$\rho_f$(\EF)}

\newcommand{\speck}{$\rho({\textbf k},\omega)$}

\newcommand{\specfk}{$\rho_f({\textbf k},\omega)$}

\newcommand{\kay}{$\textbf k$}

\newcommand{\kayperp}{k$_\perp$}

\newcommand{\kaypar}{$\textbf k_\parallel$}

\newcommand{\om}{$\omega$}

\newcommand{\EF}{$E_{\rm F}$}

\newcommand{\fpes}{f$^{1}\rightarrow $f$^{0}$}

\newcommand{\fbis}{f$^{1}\rightarrow $f$^{2}$}

\newcommand{\vo}{V$_2$O$_3$}

\newcommand{\vcro}{(V$_{1-x}$Cr$_x$)$_2$O$_3$}

\newcommand{\vtio}{(V$_{1-x}$Ti$_x$)$_2$O$_3$}

\title{The Kondo Resonance in Electron Spectroscopy}

\author{\textsc{J. W. Allen}$^{1}$\thanks{E-mail address:  jwallen@umich.edu}}

\inst{$^{1}$Randall Laboratory, University of Michigan, Ann Arbor, MI 48109-1120 USA}

\abst{The Kondo resonance is the spectral manifestation of the Kondo properties of the impurity Anderson model, and also plays a central role in the dynamical mean-field theory (DMFT) for correlated electron lattice systems.  This article presents an overview of electron spectroscopy studies of the resonance for the 4f electrons of cerium compounds, and for the 3d electrons of \vo, including beginning efforts at using angle resolved photoemission to determine the \kay-dependence of the resonance.  The overview includes the comparison and analysis of spectroscopy data with theoretical spectra as calculated for the impurity model and as obtained by DMFT, and the Kondo volume collapse calculation of the cerium \alf-\gam\ phase transition boundary, with its spectroscopic underpinnings.}

\kword{Kondo resonance, cerium, \vo, 4f electrons, 3d electrons, photoemission spectroscopy, inverse photoemission spectroscopy, Kondo volume collapse, cerium \alf-\gam\ transition}

\begin{document}
\maketitle

\section{Introduction} 

The Kondo or Suhl-Abrikosov resonance is the primary spectral manifestation of the Kondo properties of the impurity Anderson model\cite{Anderson}.  For this article the name "Kondo" is particularly appropriate and will be used henceforth.  As described further in Section 2,  the resonance is a Fermi energy (\EF) peak in the single particle local orbital spectral function \specf, i.e. the spectrum to remove or add a local orbital electron.  Over the last 25 years the status of the resonance has evolved from being experimentally un-observed and theoretically speculative\cite{Willkins} to being an accepted spectroscopic phenomenon\cite{AllenIsakson, KSAexperiment} with a rigorous theoretical description\cite{KSAtheoryReview}.  Further, the resonance has acquired a more general import because it is the central component of a new general many-body theoretical method for describing correlated electrons in lattice systems, known as dynamical mean-field theory (DMFT) \cite{DMFToverview1, DMFToverview2, LDAplusDMFT1, LDAplusDMFT2}.  This article describes the essential results of experimental studies of the resonance using photoemission spectroscopy (PES) and inverse photoemission spectroscopy (IPES).  As summarized in Section 3, PES and IPES are the spectroscopic methods whereby the spectra to remove and add electrons, respectively, can be measured. IPES is also known as Bremsstrahlung isochromat spectroscopy (BIS) when performed with X-ray photons.   

This article focuses on the 4f electrons in Ce materials and the 3d electrons of \vo\ to illustrate the evolving importance of the resonance for the electronic structures of correlated electron solids.  Ce and its compounds have long served as paradigms for research on mixed valence, the Kondo effect and the interplay of magnetism and superconductivity\cite{PairBreakingMaple}.  In solids, possible valence states for cerium atoms are 3+ and 4+, corresponding to the 4f$^1$ or 4f$^0$ configurations, respectively.  An early and very influential view of cerium materials was that their many anomalous magnetic, electrical and thermal properties involved a transition, gradual or abrupt, between the two valence states.  Already in the 1940's the loss of magnetic moments and the 15$\%$ volume contraction in the \gam\ to \alf\ transition\cite{CeTransBridgman, CeTransReview} of elemental Ce was attributed by Pauling\cite{promotion1} and Zachiariasen\cite{promotion2} to an abrupt transition from 3+ to 4+.  Similarly an absence of magnetic moments in certain "\alf-like" Ce compounds was taken to signal the 4+ valence state, and variations with pressure, or from compound to compound, of the depairing strength of cerium impurities in superconductors could also be viewed as a valence variation.  

In an alternate line of thought it was also realized that Kondo physics could account for the variation of Ce depairing strengths\cite{PairBreakingMaple, PairBreakingMullerHartman, PairBreakingGey} and the absence of magnetic moments\cite{Edelstein} in \alf-Ce.  As described in Sections 4 through 6, electron spectroscopy on cerium materials has made the decisive contributions to the eventual understanding that the formal Ce tetravalent state occurs only in insulating compounds such as CeO$_2$ or CeF$_4$ and that Kondo physics associated with the trivalent state provides the best starting point for understanding the novel properties of metallic cerium materials\cite{AllenIsakson, KSAexperiment}.  The Kondo view of the \gam\ to \alf\ transition is called the Kondo volume collapse (KVC).  Two versions can be distinguished, one\cite{ KVCLavagna1, KVCLavagna2} based on the Kondo Hamiltonian and the other\cite{KVC1, KVC2} rooted in the impurity Anderson Hamiltonian and its multiple energy scales, described in Section 2.  The latter version of the KVC can be connected to the larger spectroscopic view and is presented in Section 5.
  
\vo\ is an equally important paradigm of another fundamental phenomenon of condensed matter physics, the Mott metal to insulator transition\cite{Mott, Gebhard}.  With applied pressure or doping, as in \vcro\ and \vtio, this system displays a phase diagram with paramagnetic metallic (PM), paramagnetic insulating (PI) and antiferromagnetic insulating (AFI) phases.  The PM to PI phase boundary was identified\cite{V2O3FirstPRL,V2O3PRB} as a Mott transition within the framework of the one-band Hubbard model\cite{Hubbard}.  There is an enormous literature\cite{Gebhard} assessing this interpretation.  The most recent developments are a demonstration of the need for realism\cite{Ezhov, Parkprb} beyond that of the one-band model and the application\cite{Held} to \vo\ of DMFT combined with the local density approximation (LDA) of band theory\cite{LDAplusDMFT1,LDAplusDMFT2}.  Although the physics of the impurity Anderson and Hubbard models would seem to be quite different, a surprising twist of DMFT has been the implication\cite{DMFToverview2} that Kondo physics has great relevance for the Mott transition, and that an associated Kondo-like peak is expected\cite{Held} in the single particle spectrum of the PM phase.  PES experiments\cite{MoPMphase} observing this peak are described in Section 7

DMFT provides a new context for assessing another model of the Ce \gam\ to \alf\ transition, a view alternate to the KVC model but nonetheless within the framework of a 3+ valence state, that the Ce 4f electrons undergo a Mott transition\cite{CeMott1}.  When Kondo physics is included in the theory of the Mott transition, as in DMFT, the distinction between the KVC and the Mott views of the \gam\ to \alf\ transition begins to blur\cite{CeDMFT}.  However, the original "Mott transition" picture of the \gam\ to \alf\ transition, as defined by calculations that were done\cite{CeMott2, CeMott3} in its implementation, involves physics quite different from that of either the KVC model or the modern DMFT view of the Mott transition, precisely because both of the latter two contain essential Kondo physics.  Yet another perspective on the relation of the two pictures, presented at the outset of Section 5, concerns the temperature dependence of the Fermi surface volume of a Kondo system, as discussed in Sections 2 and 6.

\section{Important Theory Results}

This section summarizes theoretical results that provide a context for viewing the experimental findings presented in Sections 4 through 7.

\subsection{Impurity Anderson model}

The degenerate impurity Anderson model provides an elegant example of typical many-body physics, an emergent low energy scale controlled by virtual high energy scale fluctuations, with an exponential dependence of the former on the latter.  The low energy scale is the Kondo temperature, \kB \TK, characterizing local orbital magnetic moment fluctuations.  The high energy scales are for charge fluctuations characterized by the model ingredients, an \Nf-fold degenerate local orbital with binding energy $-$\epsf\ relative to the conduction electron \EF, the local orbital Coulomb repulsion \Uff, and an energy ($\varepsilon$) dependent hybridization \VE\ between the local orbital and the conduction electron states. In the discussion below the hybridization is characterized by \DE=$\pi$\VE$^{2}$ and the model definition of \VE$^{2}$ is such as to give it units of energy.  Internal energy level structure that splits the local orbital degeneracy, such as a spin-orbit splitting \DLS, or crystal field splittings, is another important ingredient.  If \Uff\ is very large and \DE\ is constant, \kB \TK=\EF\ exp[$-$1/J] where the Kondo coupling constant J = \Nf $\Delta$/$\pi$\epsf.  

The local orbital single particle spectral function is truly beautiful for displaying all of these interlocking energy scales in a single spectrum.  It was first calculated by Gunnarsson and Sch\"onhammer (GS) using an inverse degeneracy method for T=0 applied to the case of Ce materials\cite{KSAtheory1, KSAtheory2, KSAtheory3}.  The GS formulation, which has been reviewed\cite{KSAtheoryReview} in detail and also presented pedagogically\cite{AllenPed}, was enormously important for Ce spectroscopy because it was made sufficiently realistic to allow spectra to be analyzed and to be correlated with low energy Kondo properties.  This realism was extended to calculating \DE\ within the LDA\cite{KSAtheoryLDA1, KSAtheoryLDA2, KSAtheoryLDA3}.  The use of the GS method to analyze data has been reviewed\cite{KSAexperiment} in detail and examples are given in Sections 4 and 5 below.  Later calculations of the resonance using various different methods and employing the "non-crossing approximation" (NCA) \cite{NCA1, NCA2, NCA3}, were important for providing the T-dependence of the resonance, but were typically for less realistic models, e.g. a simplified \DE\ and infinite \Uff.  The T-dependence\cite{KSAtheoryReview, NCA3} of the spectrum is described further in Section 4 below and also in the discussion of \vo\ in Section 7.

The essential features of the T=0 single particle spectrum are as follows.  Consider Ce with \Nf $=$ 14 and occupation \nf $=$ 1 and suppose initially that \VE $=$ 0.  Then the spectrum consists of an \fpes\ sharp ionization peak \epsf\ below \EF\ and an \fbis\ sharp affinity peak \Uff $-$ \epsf\ above \EF.  The peaks' spectral weights are \nf $=$ 1 and (\Nf\ $-$ \nf) $=$ 13, respectively.  The ground state is magnetic and the local orbital magnetic susceptibility is a Curie law.  For any non-zero V the ground state and the spectrum are fundamentally different.  The ground state has f$^0$ and f$^2$ configurations admixed so as to stabilize a singlet below the lowest energy magnetic f$^1$ configuration.  If \Uff\ is very large, i.e. if the admixing of f$^2$ configurations can be neglected, the ground state binding energy relative to that for V=0 is simply \kB \TK.  Of great importance for this article, the spectrum develops a third peak around \EF, the Kondo resonance.  The peak maximum is nominally \kB \TK\ above \EF\ and it extends to nominally \kB \TK\ below \EF.  The ionization and affinity peaks shift somewhat and broaden to connect continuously to the resonance.  Internal energy level structure, e.g. a spin-orbit splitting, induces "sideband" peaks on the resonance, e.g. at \DLS.

The single particle spectrum provides a fine example of quasi-particle physics and the Kondo resonance can be understood as the quasi-particle of the Anderson impurity model.   The existence of the resonance can be deduced from the many-body Friedel sum rule\cite{LangerAmbegaokar, Langreth}, from which it can be shown\cite{Langreth} that at T=0 the \EF-value of the local orbital spectral function \specfEF\ depends only on \nf\ for any values of \Uff, \epsf\ and $\Delta \neq$ 0. For non-zero $\Delta$ and \Uff $=$ 0 it is clear that the ground state is non-magnetic and that the local orbital spectrum is an \Nf-fold degenerate virtual bound state cut by the Fermi energy to define a certain \nf.  One then turns on \Uff, adjusts \epsf\ to maintain \nf, and asserts that the ground state continues to be non-magnetic.  On the one hand, spectral weight must be pushed out of the virtual bound state to new PES and BIS peaks at $-$\epsf\ and \Uff $-$ \epsf, respectively, but on the other hand \specfEF\ is fixed and so an \EF\ peak of diminished spectral weight, the Kondo resonance, must remain.  Before a rigorous calculation of \specf\ was available, this logic was a primary theoretical evidence\cite{Willkins} for the existence of the resonance and indeed, the GS calculation satisfies the Friedel sum rule.

The most basic features of the integrated spectral weights of the various peaks are well illustrated by results for large \Uff\ and no internal energy level structure, i.e. no sideband peaks.  The weight of the \fbis\ feature is the product of the probability of f$^1$ in the ground state multiplied by the number of possible final states, i.e. (\nf)(\Nf$-$1).  The BIS part of the resonance corresponds to {f$^{0}\rightarrow$f$^{1}$} transitions and so by similar reasoning it has weight (1$-$\nf)(\Nf).  These sum to (\Nf$-$\nf) for the total BIS weight, as they must.  The total spectral weight of the PES part is \nf\ so that the total spectral weight of the combined PES/BIS spectrum is \Nf, as must also be so.  The weight of the PES portion in the near \EF\ resonance can be estimated as the product of its width \kB \TK\ and \specfEF\ = $\pi$\nf$^2$/(\Nf $\Delta$), as fixed by the Friedel sum rule for the simple case of constant \DE.  Using the simple relation\cite{KSAtheory2} (1$-$\nf)/\nf\ = $\pi$\nf \kB \TK /\Nf $\Delta$, one finds thereby a weight (1$-$\nf)\nf.  The small quantity (1 $-$ \nf) is seen to be significant for giving the ratio of the spin to the charge energy scales and for giving the fractional weights of both the PES and BIS parts of the resonance.

Resonance sideband peaks, e.g. at \DLS, provide easily observable PES spectral weight near \EF\ even for very small (1$-$\nf) and \TK.  Therefore the total weight of the sidebands plus the \EF\ resonance is no longer just (1$-$\nf)\nf.  To understand why, it is helpful to view the Kondo resonance as a kind of relaxation or screening process in which hybridization allows the f-hole to be filled by a conduction electron from near the Fermi level, thereby restoring the ground state occupation of the local orbital, giving a much reduced one-hole energy, and producing a peak in \specf\ within \TK\ of \EF. For non-zero \DLS, the filling of the f-hole need not yield the lowest energy state, but can, with some probability, also lead to the spin-orbit excited state and hence produce a peak in the spectrum at \DLS\ below \EF. The internal energy level splitting \DLS\ also decreases the ground state degeneracy, e.g. from 14 to 6 for Ce$^{3+}$, which reduces J proportionately, other things being equal.   The value of \TK\ is then reduced exponentially and thereby the weight in the resonance peak just at \EF\ is greatly reduced.  Although there is no rigorous proof, it is plausible and is found\cite{AllenAdvPhys} numerically to be the case, that the total weight of the \EF\ peak plus its sidebands is roughly conserved as \DLS\ becomes non-zero.  Thus one can think of the large sideband weight as being the large \EF\ resonance weight that would occur if \DLS\ were zero and \TK\ were large.

\subsection{Lattice Aspects}

\subsubsection{Dense Impurity Ansatz}

The Ce materials actually studied are usually concentrated lattice systems and so should be described by the lattice Anderson model.  The impurity model has often been applied to such systems, a scheme that is sometimes called the "dense impurity ansatz."  This ansatz can be stated more precisely.  For the lattice system the spectral function should depend on \kay, the crystal momentum, i.e. one has \specfk.  A local spectral function can then be defined by summing over all \kay, i.e. \specfLOC $= \sum _{\textbf k} $ \specfk.  The ansatz amounts to the assumption that the impurity spectral function provides an approximation to the local one, i.e. \specfLOC $\approx$ \specf.  The success of the impurity model for describing angle-integrated Ce 4f PES spectra is empirical evidence for this plausible assumption.  Thus the observation of \kay-dependence in f-electron spectra, described in Section 6, does not imply a failure of the density impurity ansatz for \specfLOC, as has sometimes\cite{Arko2} been supposed. 

A theoretical rationale can also be offered for the ansatz.  If the self-energy of the Green's function\cite{Hedin} that underlies \specfk\ is \kay-independent, one can show\cite{NoGo} that \specfLOC\ satisfies a sum rule exactly like the Friedel sum rule, i.e. \specfLOCEF $=$ \specfNONEF, where \specfNON\ is the spectral function for a non-interacting system.  Further\cite{KSAtheoryLDA2}, if \DE\ is obtained from some one-electron approximation for the lattice, e.g. the LDA, the impurity calculation then includes f-electrons on all sites, with many-body effects being treated explicitly on one site, while the other sites are treated at the level of the LDA.  In this case, if \Uff\ is set to zero, \specf\ will be the LDA \specfLOC, i.e. the f-density of states of a lattice calculation.  Thus if the LDA is used to find \DE\ and is taken to be the non-interacting system, and if the resulting self energy were \kay-independent, then the dense impurity ansatz would be exact just at \EF, i.e. \specfEF\ $=$ \specfLOCEF.  It is then tempting to speculate that if the self energy is \kay-independent, there might be some way to formulate a more sophisticated "dense impurity ansatz" that would be exact for all $\omega$, i.e. \specf\ $=$ \specfLOC.  In fact, this is a heuristic, albeit imprecise and narrow, way of viewing DMFT, the topic of the next sub-section.  
 
\subsubsection{DMFT and the Mott Transition}

DMFT is a self-consistent many-body method for studying quantum-mechanical lattice problems with local interactions, where a lattice site is assumed to be surrounded by very many, in fact, infinitely many, nearest neighbors, in analogy with classical mean-field theories\cite{DMFToverview1}. While spatial fluctuations are neglected in DMFT the quantum dynamics of the many-body problem, which is of prime importance for strongly correlated electrons, is completely included. This scheme can be shown\cite{DMFTinfdim} to become exact in the limit of the coordination number going to infinity.  It may be formulated as a mapping\cite{DMFTmapping1, DMFTmapping2} of the lattice problem onto an effective Anderson impurity model coupled self-consistently to an effective conduction band bath.   In application to the Hubbard model DMFT is the best description that can be made by using a \kay-independent self energy.

For describing the Mott metal to insulator transition within the framework of the Hubbard model the self-consistency step is crucial because it enables a description\cite{DMFToverview2, DMFTmapping1} of the transition in which the spectral weight of the metal phase quasi-particle decreases with increasing ratio of Coulomb energy to bandwidth until the weight eventually goes to zero in the insulating phase.  Very near the transition on the metal side one might think of the tiny remaining quasi-particle weight as a "bootstrapped" Kondo resonance, in which the \EF\ resonances of all the sites combine to form the conduction electron bath needed by any one site to form its resonance. 

A new development that enables realistic descriptions of correlated systems within a quasi-particle framework is the mating of the DMFT with LDA band theory\cite{LDAplusDMFT1, LDAplusDMFT2}.  LDA+DMFT is well on the way to having the workhorse stature for solid state electronic calculations that is now enjoyed by the LDA.  Section 7 describes PES spectra for the 3d electrons of the PM phase of \vo\ and gives a comparison\cite{MoPMphase} to the local spectral function calculated within the DMFT+LDA.   

\subsubsection{Lattice Anderson Model}

Relative to the impurity model, the obvious new spectral aspect of the Anderson lattice model is the \kay-dependence of \specfk\ and, within Fermi liquid theory, the implied existence of a Fermi surface (FS).  The Luttinger counting theorem\cite{Luttinger} is the general T=0 Fermi energy sum rule for a lattice system, analogous to the Friedel sum rule\cite{Langreth} for the impurity Anderson model.  The content of the theorem is that the volume enclosed by the Fermi surface of an interacting system is the same as that of a non-interacting reference system chosen to have the same ground state symmetry, i.e. a phase transition should not occur as Coulomb interactions are turned on.  In essence, the number and quantum labeling of the electrons is unchanged by the interactions, and so the counting of them remains the same.  An important implication\cite{LuttingerMartin} is that if the magnetic moments of, e.g., the f-electrons are quenched, by any mechanism, including the Kondo effect, then the f-electrons must be counted in the Fermi surface volume.  FS studies\cite{felecdHvA} using magneto-oscillatory techniques such as the de Hass van Alphen (dHvA) effect have verified that this is so for heavy Fermion f-electron systems.  An important theoretical conjecture\cite{LuttKondoHiT1, LuttKondoHiT2} is that if the temperature is raised to well above \TK, so that the f-moments are no longer quenched, the f-electrons should be excluded from the FS, i.e. the FS volume should count only the conduction electrons.  As presented in Section 6, high T ARPES spectra support the correctness of this conjecture, leading to a further perspective on the relation of the KVC and Mott models for the \gam\ to \alf\ Ce transition, set forth at the beginning of Section 5. 

The theory of \specfk\ is evolving.  The standard approach to the FS simply includes the f-electrons in an LDA band structure calculation\cite{CeRuSiRenBands}.   A more sophisticated band structure approach includes the f-electrons but renormalizes their \EF\ scattering phase shifts to be Kondo-like, a procedure that gives enhanced masses and often makes only minor perturbations on the LDA FS\cite{RenBands}.  This method is not designed to provide a spectral function that shows the various energy scales and spectral weights and their interplay.  Treatments\cite{latAnd1, latAnd2, latAndJarrell} of a simplified Anderson lattice Hamiltonian have led to a renormalized f-d mixing model that is summarized and compared to ARPES spectra in Section 6, although such simplified models are limited for comparison to ARPES and FS data because they lack material specific realism.  

The "exhaustion" problem\cite{Nozieres}, raised for the s-state Kondo lattice, is an interesting issue that bears on the dense impurity ansatz.  It is argued that the energy scale entering the specific heat should decrease with increasing concentration to reflect a lack of conduction electrons to Kondo screen the magnetic moments of all the sites.  This lack is understood to arise because only a small fraction of the conduction electrons, those within \TK\ of \EF, have energies low enough to be involved in the screening.  Numerical simulations\cite{latAndJarrell} of the \Nf\ = 2 Anderson lattice have given support to the exhaustion idea but experimental evidence is either lacking or negative.  For example, in the system La$_{1-x}$Ce$_x$Al$_2$ the value of \TK\ deduced from the specific heat increases with x because the volume decreases as the larger La$^{3+}$ ion is replaced by the smaller Ce$^{3+}$ ion, giving an increased hybridization\cite{CeDilTK}.  As presented in Section 4, PES spectra for small and large x are consistent with this variation of \TK.  Thus other factors, such as the large degeneracy, the underlying charge degrees of freedom of the f-electrons, or the volume change outweigh any effects due to exhaustion in this system.  The exhaustion issue has also motivated questions \cite{Nozieres} and answers\cite{KotliarOnExhaustion} concerning the application of DMFT to the Mott transition. 

DMFT+LDA is the best current possibility for realistic spectral calculations in the spirit of the lattice Anderson model.  Relative to the Hubbard model, the new feature of the lattice Anderson model is the presence of non-interacting conduction electrons to which the strongly correlated electrons are hybridized.  Based on the empirical success of the dense impurity ansatz, as described in Sections 4 and 5, one can speculate that the self-consistency step of DMFT is less essential for determining \specfLOC\ in this case than for the Hubbard model.  It may be that by adjusting \epsf, \Uff\ and the overall magnitude of the LDA \DE, as is typically done, one catches something of the self-consistency effects.  However the effect of hybridization on the conduction electrons for a lattice system cannot be obtained in the impurity model and has recently been studied using DMFT in the context of the Ce \gam\ to \alf\ transition \cite{KVCKotliar}.  Further, the \kay-space distribution of f spectral weight is now being studied by ARPES, as described in Section 6.  One can show\cite{NoGo} that if the self-energy is \kay-independent, not only is \specfLOCEF $=$ \specfNONEF, as mentioned above, but also the FS shape is preserved.  Thus DMFT+LDA should be capable of providing valuable guidance for analyzing ARPES studies of lattice Anderson systems, but such comparisons of theory and experiment remain to be done.
   
\section{Electron Spectroscopy Basics}

In PES photons incident on a sample excite photoelectrons which are detected and analyzed for their kinetic energies, whereas in IPES the reverse experiment is performed.  Descriptions of PES and IPES are given in detail in books and articles\cite{PESgen, ARPESgen1, ARPESgen2}.  Most of the present article concerns PES spectra and so the content of this section is limited to a summary of some basics of PES that are relevant to the data presented in this article. The basics of IPES\cite{BIS, IPES} are similar, although there is one major conceptual and practical difference, that the IPES cross-section is much smaller than for PES, other things being equal, because\cite{IPES} of the difference in the final state phase spaces for photons and electrons.

In angle-resolved PES (ARPES) the sample is an oriented single crystal and the emission angles relative to the crystal axes are also analyzed.  If the photoemission event can be described within the sudden approximation, the resulting ARPES spectrum can be interpreted\cite{PEStheory} as the product of the photoemission cross-section and the single-particle spectral function for electron removal, \speck\, where \om\ is the electron binding energy deduced from the difference between the photon energy \hnu\ and the kinetic energy, usually measured relative to the Fermi energy \EF, or for non-zero temperature T, the chemical potential \moo.  For fixed \hnu, the variation of the detection angles moves \kay\ on a spherical \kay-space surface and changing \hnu\ varies the radius of the spherical surface.  The component of \kay\ parallel to the measured surface, \kaypar, is preserved as the photoelectron leaves the solid so the photohole \kaypar\ is easily determined.  However \kayperp, the perpendicular component of \kay, changes as the electron traverses the sample surface potential, so the surface potential must be modeled in some way to deduce the photohole \kayperp.

The photoelectron lifetime can have an important effect on the spectra\cite{KPerpLifetime}.  It induces uncertainty in \kayperp, which complicates ARPES studies of three-dimensional materials.  It also induces an extra and unwanted contribution to the photohole linewidth, which is proportional\cite{PEStheory} to the perpendicular photohole velocity.  Thus the undesirable effects of the photoelectron lifetime are  absent for a system with dimensionality strictly two or less.

Improved instrumentation has enabled a new ARPES technique in which the analyzer kinetic energy is fixed and the analyzer angles are varied to provide an intensity distribution for a fixed photohole energy.  When the energy is chosen to give the Fermi energy, the resulting intensity distribution can be interpreted as a FS map\cite{FSMap}. Obtaining FS maps for fully three dimensional materials is challenging because of the need to determine \kayperp\ and the possible loss of \kayperp\ resolution due to the photoelectron lifetime.  Examples of FS maps for three-dimensional systems are shown in Section 6. 

Determining \speck\ from the measured spectrum also requires disentangling the effects of the photoemission cross-section, which varies with \hnu\ in ways related to the underlying atomic characters\cite{Yeh} of the states being probed, with \kay\ in ways that reflect the underlying band structure\cite{Lindroos}, and with photon polarization in ways that reflect both.  When these dependences are sufficiently well understood, they can be used to advantage to enhance emission from states of particular interest.  A very useful photon energy dependence is that of element and orbital specific cross-section resonances that occur at specific core level absorption edges.  As an example relevant for this paper, cerium 4f emission can be greatly enhanced for photon energies at the Ce 3d and 4d edges.  This technique is known as resonant PES (RESPES) \cite{RESPES}.  Similar resonant enhancements in Ce 4f BIS have also been demonstrated\cite{RESBIS3dedge, TdepCeRESBIS, RESBIS4dedge}.

Photoemission is surface sensitive because elastic escape depths\cite{EscapeDepth} for photoelectrons are often no greater than one or two lattice constants.  Even if a measured surface is free of contaminants the reduced atomic coordination on the surface can cause the surface electronic structure to be different from that of the bulk.  As discussed in Sections 4, 5 and 7 in this article, such surface effects are especially important for the strongly correlated systems discussed in this paper.  One attack on this problem is to observe the effect of changing the probe depth.  The probe depth can be varied by changing \hnu\ because the elastic escape depth changes with kinetic energy and hence with \hnu.  The dependence of escape depth on energy tends to be a U-shaped curve with a minimum in the range 50 eV to 100 eV and so enhanced bulk sensitivity can be achieved for very low photon energies, accessible with lasers, and with higher photon energies, nominally greater than 300 eV.  Recently the resolution of PES at these higher photon energies has been greatly improved\cite{BL25SU}, leading to a considerable increase of effort on this issue.  For fixed \hnu\ the effective probe depth perpendicular to the surface is greatest for normal emission and is reduced as the detection angle varies away from normal emission.

Much of this article involves comparison of PES (i.e. angle integrated) spectra with theoretical calculations of \spec, i.e. the full \kay-sum of \speck.  From the discussion above one sees that the PES spectrum for fixed \hnu\ is only a partial \kay-sum and should further be averaged over a range of \hnu\ large enough to traverse a full Brillouin zone in order to obtain \spec.  However if the smearing of the perpendicular component of \kay\ due to the photoelectron lifetime is as large as a full Brillouin zone, which is possible, such additional averaging is not necessary.  The discussion above shows that the variation of \hnu\ also may vary the cross-section, especially in the case of RESPES, which introduces a complication that is difficult to account for quantitatively.  Thus it has been most common to compare a PES spectrum for fixed \hnu\ with a theoretical \spec, although the need to average over a range of \hnu\ was addressed in recent work\cite{MoPMphase} on \vo.

\section{Angle Integrated PES for Cerium Materials}

This section presents an overview of the Kondo resonance as seen in angle integrated electron spectroscopy of the 4f electrons of Ce materials.  From this spectroscopic viewpoint the current Kondo picture of Ce started in 1979 with the use of RESPES to determine\cite{FirstCe1} that \epsf\ of Ce is $\approx$ 2eV, confirming an important theoretical prediction\cite{CeMott1} that motivated the Mott transition model of the \gam\ to \alf\ transition.   In 1981 estimates of Kondo temperatures from the widths and energies of Ce 4f ionization peaks were made\cite{FirstCe2, FirstCe3} within the framework of the non-degenerate impurity Anderson model, and in 1982 RESPES spectra clearly showed\cite{firstCeRu2RESPES} a large 4f occupancy in superconducting Ce compounds previously thought to be tetravalent.  These early efforts served as an experimental motivation for the KVC model\cite{KVC1} based on the Anderson Hamiltonian, also put forth in 1982.  The subsequent development in 1983 of the GS theory\cite{KSAtheory1} of the resonance for the degenerate impurity Anderson model enabled the resonance to be identified\cite{KRFirst} in combined 4f RESPES and BIS spectra.  An important additional component of the GS work was a theory for the Ce 3d core level PES spectrum, which contains structure\cite{Ce3dXPS} than can be analyzed\cite{CeAnalyze} as an additional independent source of information on the values of Anderson model parameters.  

Fig. 1 shows spectra from an overview article\cite{AllenAdvPhys} on this early work, along with theoretical PES and BIS spectra calculated using the GS method.  The experimental PES spectra were obtained using the RESPES technique at the Ce 4d edge around 122 eV, with an overall energy resolution of 0.4 eV, and the BIS spectra were obtained in a laboratory instrument, detecting 1486.6 eV photons with an overall resolution of 0.6 eV.  The basic three features of the 4f spectrum are clear, an ionization peak roughly 2 eV below \EF, an affinity peak roughly 4 eV above \EF, and an \EF\ resonance that is particularly large for the large \TK\ \alf-like material. An important additional component of this work was the use of Ce 3d core level spectra (not shown here) to further constrain the Anderson model parameters.  T=0 magnetic susceptibility values calculated using the spectroscopically derived impurity model parameters were found to be in fair agreement with experimental values.

\begin{figure}[!tb]
\begin{center}
\includegraphics[width=12cm]{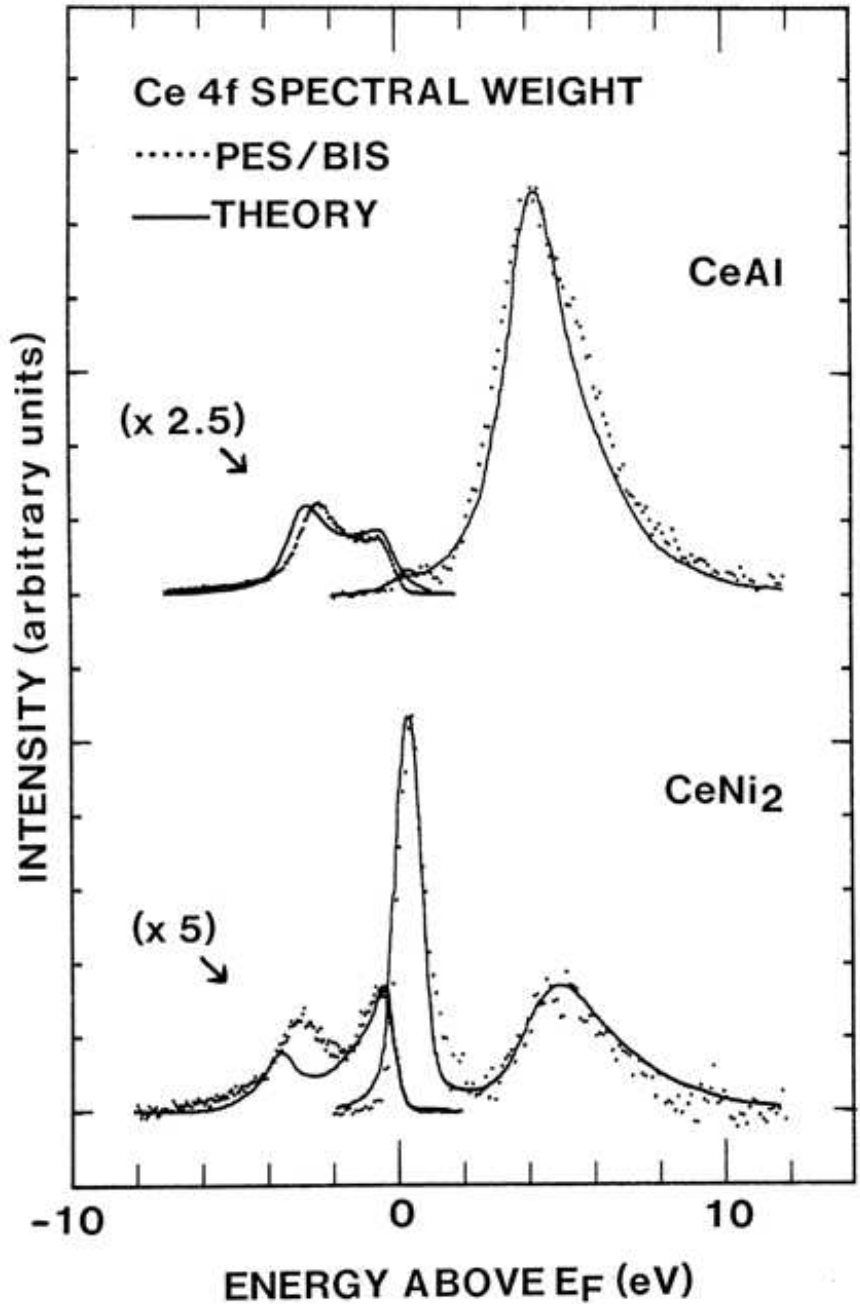}
\end{center}
\caption{First combined Ce 4f RESPES and BIS spectra of small (CeAl) and large (CeNi$_{2}$) \TK\ materials, compared to Anderson impurity model GS spectral theory, showing ionization and affinity peaks well below and well above \EF, respectively, and growth from small \TK\ to large \TK\ of Kondo resonance centered just above \EF.  (from Ref. [\cite{AllenAdvPhys}])}
\label{f1}
\end{figure}

\begin{figure}[!tb]
\begin{center}
\includegraphics[width=12cm]{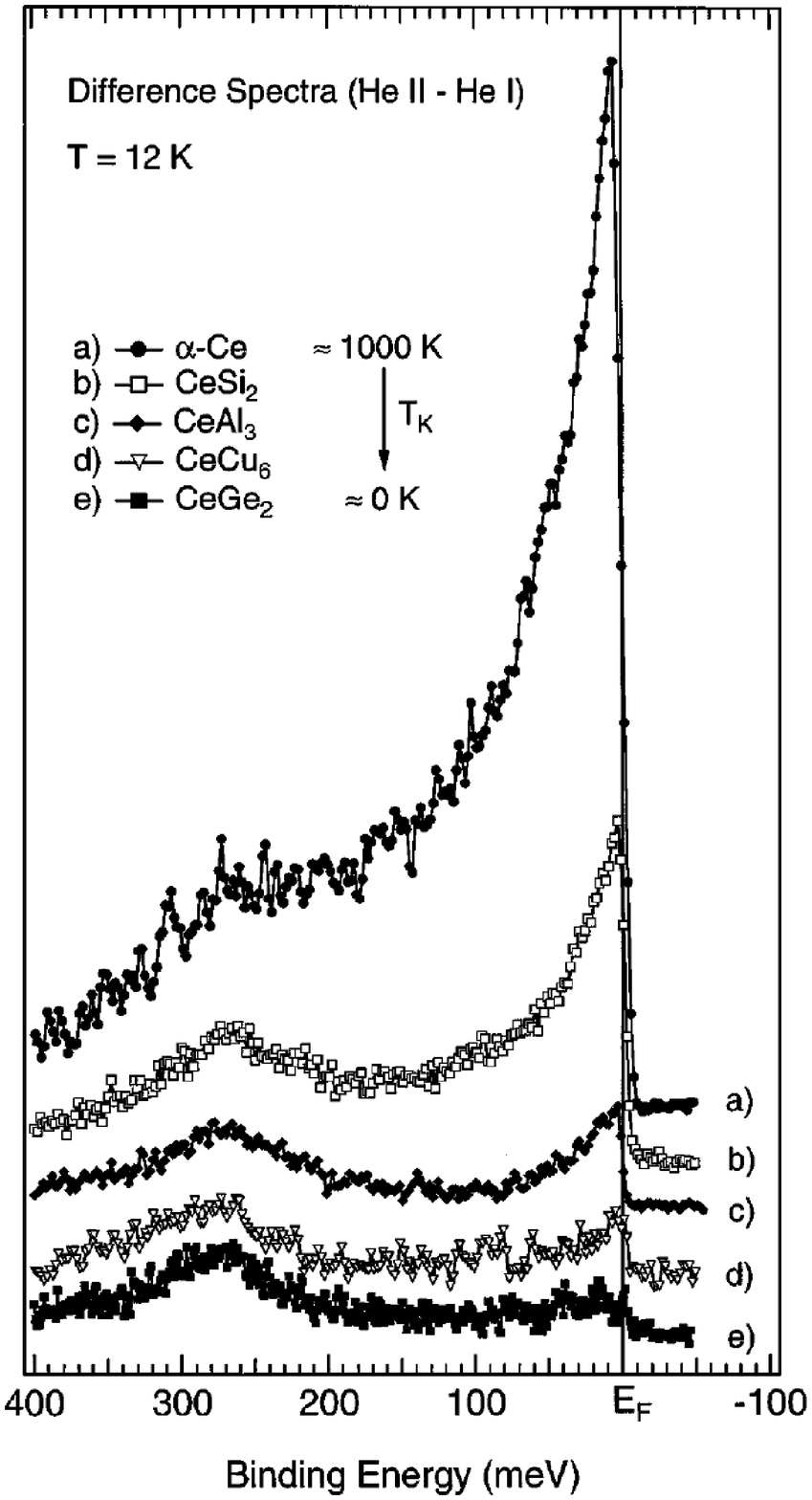}
\end{center}
\caption{ Ce 4f PES spectra with resolution 5 meV for materials with a range of \TK\ values, showing evolution with increasing \TK\ of near \EF\ and 280 meV spin-orbit sideband weights of Kondo resonance.  (from Ref. [\cite{NewestCeHiResTDep}])} 
\label{f1}
\end{figure}

A very important contribution to PES work on the resonance has been a steady improvement in resolution, beginning\cite{FirstCeHiRes} in 1985 with the first observation of the spin-orbit sideband at \DLS\ $\approx$ 280 meV below \EF\ for \gam\ and \alf\ Ce, and continued for many Ce systems\cite{BaerAdvPhys}.  The most recent work\cite{NewestCeHiResTDep, TdepCePESHufner1, TdepCePESHufner2} has a resolution as good as 5 meV, roughly a factor of 100 better than in the earliest work.  Fig. 2 shows such high resolution 4f PES spectra\cite{NewestCeHiResTDep} for a series of materials with varying values of \TK.  The 4f spectrum has been extracted using the PES cross-section variation over an \hnu-range available from a laboratory helium resonance lamp, 20 eV to 40 eV.  The spin-orbit sideband is clearly seen.   The increasing weight of the \EF\ resonance relative to that of the sideband for increasing values of \TK\ is generic. As was appreciated in early work\cite{AllenAdvPhys}, and as can be seen in Fig. 2, the sideband gives easily observable weight even for small \TK\ materials and the same is true for sidebands due to crystal field splittings.  Thus this aspect of the experimental spectra does not\cite{CommentOnArko1, CommentOnArko2} constitute evidence against the use of the impurity model, as has sometimes\cite{Arko1, ArkoReply} been claimed.  Indeed on purely conceptual grounds the presence of the sideband in the spectrum is a clear signature of the basic many-body working of the model because \DLS\ is an energy characteristic of the f$^{1}$ configuration, and hence would not be seen if the spectrum were simply of \fpes\ character. Another source of near \EF\ PES weight in addition to that of the Kondo resonance itself can be {f$^{2}\rightarrow$f$^{1}$} transitions\cite{KSAtheory3} allowed by f$^{2}$ states present in the ground state for finite \Uff.

For fixed model parameters a slow temperature dependence, nominally as ln(T/\TK) for small \TK, is predicted for the resonance. Differing behaviors of the PES and BIS parts provide an interesting manifestation of the essential asymmetry of \nf\ $\ll$ \Nf\ in the degenerate impurity model for Ce. The BIS part of the resonance is essentially of f$^{0}\rightarrow$f$^{1}$ character and hence the integrated weight reflects the thermal average (1-\nf(T)) of the initial state, which changes from its ground state value toward 1 as non-magnetic states are thermally populated for T $\gg$ \TK.  This change in \nf(T)) is fundamental.  In the GS calculation it is manifestly clear that the lowered energy of the non-magnetic ground state relative to that of magnetic states arises because of mixing between f$^{0}$ states, inherently non-magnetic, and f$^{1}$ states of non-magnetic character, a mixing which cannot occur for f$^{1}$ states of magnetic character.  This T-dependence of the BIS spectrum has been observed \cite{TdepCeBIS, TdepCeRESBIS}.

In contrast to the T-dependence of the BIS part, the integrated weight of the PES part from \EF\ down through the region of the sideband features is only weakly T-dependent, because increasing T populates magnetic states for which electron removal can lead\cite{KSAtheoryReview, KSAtheory2} to f$^{1}$ final states of magnetic character in a range near \EF.  Thus the dominant effect observed in PES is a broadening\cite{BaerAdvPhys, NewestCeHiResTDep}, for which it has been more difficult\cite{ArkoComment} but nonetheless possible\cite{TdepCePESHufner1, GarnierReply}, to experimentally demonstrate a behavior that is peculiar to the impurity Anderson model.  As described in Section 7, a similar high temperature broadening is predicted but not yet observed for the V 3d quasi-particle peak observed\cite{MoPMphase} in the PM phase of \vcro, although a closely related gap-filling effect has been observed at high temperatures for the PI phase\cite{MoPIphase}.  Section 6 takes up the question of the T-dependence that might be expected for f-electron ARPES spectra, especially in connection with the volume of the FS below and above \TK, as already mentioned in Section 2.  
    
Fig. 3 illustrates a different kind of experimental test\cite{CeDilutionKim} of the dense impurity ansatz, in the form of 20K PES data for polycrystalline samples of La$_{1-x}$Ce$_x$Al$_2$ with x=1 and 0.04.  The idea of the experiment is to compare spectra for a concentrated system and a system sufficiently dilute as to be essentially in the impurity limit. The Laves phase structure of CeAl$_2$ is well suited for this purpose because it has only four near Ce neighbors of a Ce site.  For x=0.04 a fraction (1-0.04)$^4$ = 85\% of the Ce sites are completely isolated if next-nearest neighbor Ce-Ce couplings can be neglected, so that the x=0.04 spectrum is very much dominated by single site contributions.  Obtaining the Ce 4f spectrum of such a dilute sample required high resonance contrast obtainable only by using RESPES at the Ce 3d edge, performed on beamline BL25SU\cite{BL25SU} at the SPring-8 Synchrotron.  The high contrast results from the small off-resonance cross-section at this relatively high photon energy.  The high photon energy also enhances the bulk sensitivity, as discussed further below.

\begin{figure}[!tb]
\begin{center}
\includegraphics[width=12cm]{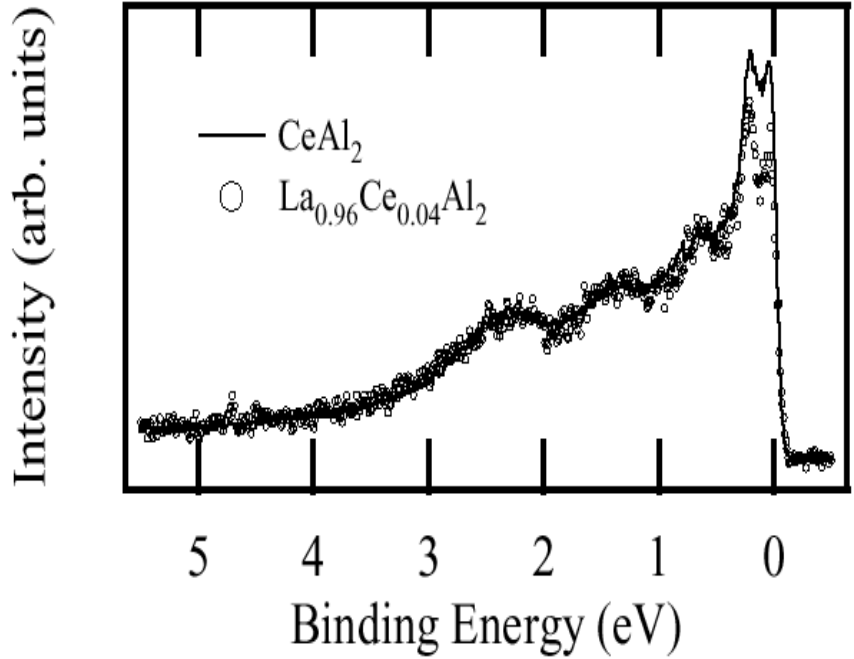}
\end{center}
\caption{Direct experimental verification of the dense impurity ansatz for a small \TK\ Ce material, as shown by nearly identical Ce 4f PES spectra of a dilute and concentrated Ce system.  Spectra were obtained by RESPES at the Ce 3d edge, with the Kondo resonance spin-orbit sideband at 280 meV resolved.  Small changes in the near \EF\ and sideband weights with dilution are as expected both from experiment on concentrated compounds (see Fig. 2) and from impurity Anderson model theory, for the known decrease of \TK\ with dilution.  (From Ref. [\cite{CeDilutionKim}])}
\label{f1}
\end{figure}

One sees in Fig. 3 that the dilute and concentrated spectra are completely identical for binding energies greater than 0.5 eV.  The multiple \fpes\ ionization peaks arise from structure in the s-p-d density of conduction electron states to which the f$^0$ final state is hybridized.  The small changes in the near \EF\ resonance and spin-orbit sideband peaks, i.e. an overall reduction of both and a greater reduction of the former relative to the latter,  are exactly as expected within the impurity theory due to the known reduction of \TK, from 4.8K for x=1 to 0.4K for x=0.04.  This decrease in \TK, deduced from the T-linear specific heat coefficient\cite{CeDilTK}, is caused by the decreased hybridization that accompanies a volume expansion as the larger La$^{3+}$ ion replaces the smaller Ce$^{3+}$ ion.  As discussed in Section 2, the decrease is opposite to expectations in the "exhaustion" picture of the s-state Kondo lattice.

Another important improvement in the quantitative analysis of angle integrated Ce spectra has come from the realization that there can be a significant surface contribution for which \TK\ is different from that of the bulk.  The origin of the effect is the reduced coordination at the surface relative to that of the bulk, because one-electron bandwidths and hybridization strengths are nominally proportional to the coordination number.  Thus the magnitude of $\Delta$ is reduced and there is a related reduction in the cohesive energy of the surface relative to that of the bulk, which leads\cite{SurfaceShiftJohansson} to an increase in the magnitude of \epsf.  The underlying sensitivity to coordination number implies\cite{SurfaceShiftJohansson} an increasing effect as the local geometry varies from a smooth surface to edge-like to corner-like, i.e. as the surface becomes rougher, and this dependence has been observed\cite{SurfaceShiftDomke} for surface shifts of \epsf. 

For Ce the changes in \epsf\ and \DE\ both act to greatly reduce \TK, which causes the observed spectrum to have less weight near \EF\ than would be expected\cite{CeTransLiu}.  Thus the effect is particularly important for large \TK\ systems.  As discussed in Section 2 the use of higher photon energy reduces PES surface sensitivity and early\cite{CeSurfaceLaubschat, SurfaceDuo} high photon energy studies showed the surface effect, although with relatively poor resolution.  Improvements in resolution\cite{BL25SU} have enabled the spin-orbit sidebands to be resolved\cite{Sekiyama}, as in the data of Fig. 3.  Before the feasibility of making high photon energy studies with good resolution, a less direct method\cite{CeTransLiu} was used to extract the bulk 4f spectra of \alf\ and \gam\ Ce, with results as illustrated in Section 5 below.  A final remark on this topic is that low photon energy spectra are not automatically precluded from being bulk-like, if the crystal structure is such that the spectra of interest come from electrons residing in a sub-surface layer.  CeRu$_2$Si$_2$, discussed in Section 6, has two cleavage planes, one of which finds the Ce atoms in a sub-surface layer.   It is found that for some cleaves, presumably ones giving this situation, the 4f spectra are essentially the same at both low and high photon energies\cite{CeSurfDenlinger}.

Another thread in the Ce PES literature is concerned with whether\cite{HiTKCe1, HiTKCe2, CeRu2two, Sekiyama} or not\cite{HiTKCe3, CeRu2one, HiTKCe4} band theory might provide a better starting point than the dense impurity ansatz for some extremely high \TK\ \alf-like materials, notably CeRh$_{3}$ or CeRu$_{2}$.  For CeRh$_{3}$ the band picture was advocated on the finding\cite{HiTKCe1, HiTKCe2} that the BIS spectrum does not show the f$^2$ peak, although other workers\cite{SurfaceDuo} have identified such a BIS peak for CeRh$_{3}$.  The most recent PES work\cite{HiTKCe4} addressing this issue utilized high resolution RESPES at the Ce 3d edge and emphasized the need to choose the resonance \hnu\ carefully to avoid a misleading Auger contribution to the Ce 4f RESPES spectrum.  The resulting 4f spectra of several large \TK\ materials, including CeRh$_3$, are described reasonably well by the dense impurity ansatz and very badly by band theory.  For CeRu$_{2}$ early uncertainty\cite{earlyCeRu2BISBaer1, earlyCeRu2BISBaer2} as to the existence of the f$^{2}$ BIS peak was resolved\cite{KRFirst} on the positive side long ago, so the usual large value of \Uff\ is operative.  Thus the BIS spectrum is clearly beyond the reach of band theory, but can be well described\cite{AllenAdvPhys} by the dense impurity ansatz, along with the low resolution 4f RESPES spectrum and the Ce 3d core level spectrum.  The higher resolution 4f RESPES spectrum obtained for CeRu$_2$ at the Ce 3d edge is anomalous\cite{Sekiyama} compared to the spectra of all other metallic Ce materials and at present merits continued experimental study.

\section{Kondo Volume Collapse and the Kondo Resonance in the Ce \gam\ to \alf\ Transition}

The experimental \gam\ to \alf\ Ce phase boundary is shown in Fig. 4, with the magnetic $\beta$-phase omitted for simplicity.  The box around the critical point shows experimental uncertainty that is actually characteristic of the entire phase boundary due to hysteretic effects\cite{CeTransReview}.  The basic idea of the KVC model\cite{KVC1} is that in the transition the system switches its Kondo temperature from \TK\ $\ll$ T in the \gam-phase to \TK\ $\gg$ T  in the \alf-phase.  The switch occurs by means of the large \alf-phase volume contraction, which increases the hybridization.  \TK\ varies rapidly with the hybridization and hence increases strongly.  The motivation for the volume collapse is the gain of the binding energy of the large \TK, which more than offsets the increase in lattice energy arising from the bulk modulus associated with all the non-f electrons.  The essential motivation for the small \TK\ of the \gam-phase is the gain of the entropy of the Ce magnetic moments, which decreases the free energy more than enough to compensate the loss of Kondo binding energy.  Section 6 provides experimental evidence for the proposal\cite{LuttKondoHiT1, LuttKondoHiT2} that the Ce 4f electron is excluded from the Fermi surface as T varies from  T $\ll$ \TK\ to  T $\gg$  \TK.  In the KVC model the switching of \TK\ then implies a similar 4f electron FS exclusion in the transition from  \alf\ to \gam.  This change in FS volume could well be called a Mott transition of the 4f electrons, although the crucial role of the Kondo effect in controlling the energy scale and entropy of the transition was not a part of the original\cite{CeMott1} "Mott transition" proposal.

\begin{figure}[!tb]
\begin{center}
\includegraphics[width=12cm]{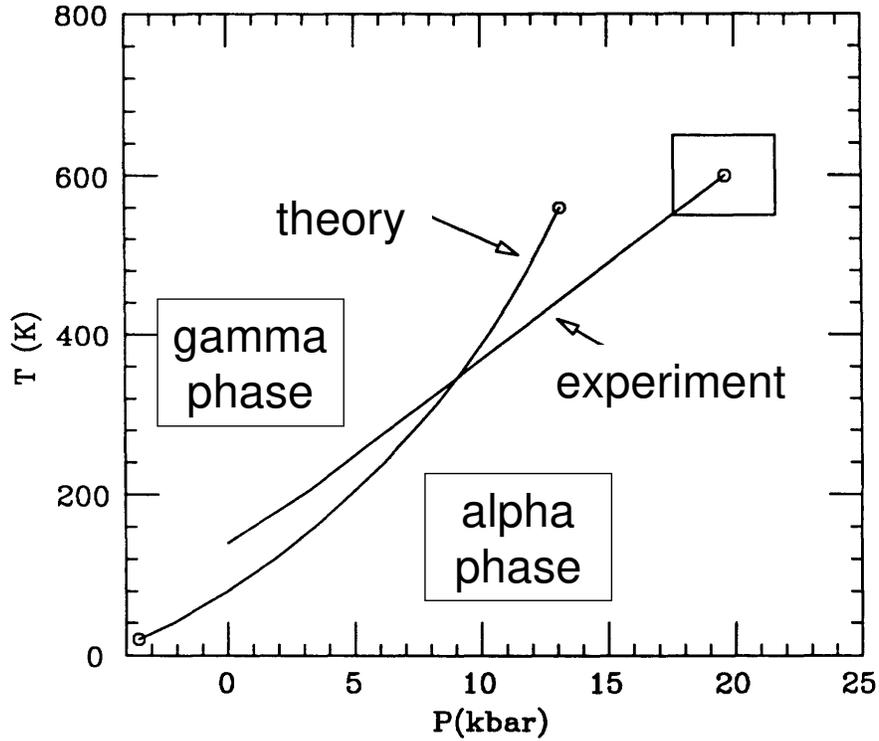}
\end{center}
\caption{Comparison between spectroscopy based KVC model calculation and experiment\cite{CeTransReview} for the Ce \alf-\gam\ phase boundary.  (from Ref. [\cite{KVC2}]).  The magnetic $\beta$-phase boundary is omitted for simplicity.    The box around the experimental critical point indicates experimental uncertainty due to hysteresis in the first order transition.  The second critical point at negative pressure has been observed, as discussed in the +text.}
\label{f1}
\end{figure}

\begin{figure}[!tb]
\begin{center}
\includegraphics[width=12cm]{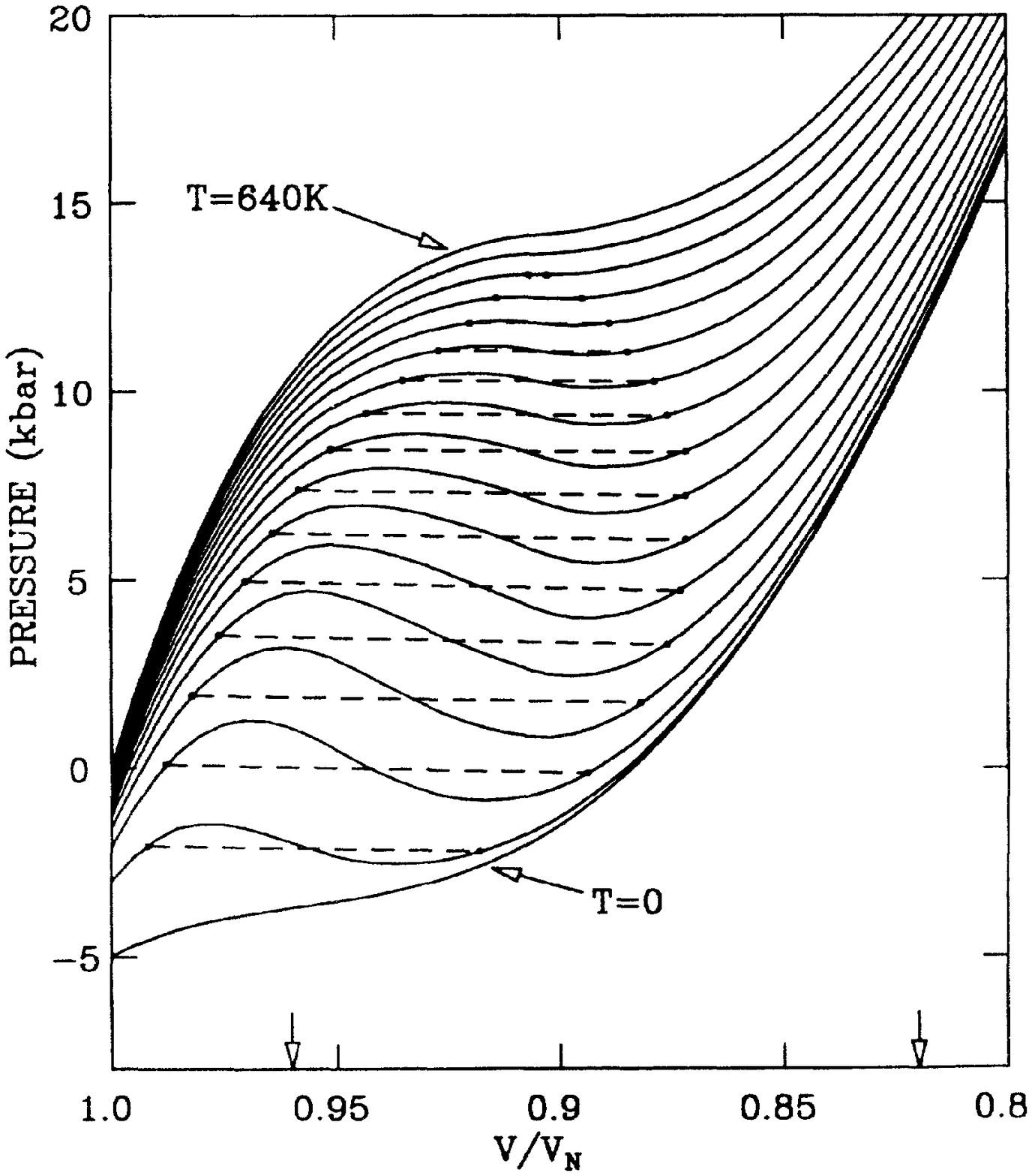}
\end{center}
\caption{P-V isotherms for T from 0 to 640K with equal increments of 40K.  Experimental equilibrium volumes of the \alf\ and \gam\ phases at T=300K are marked with arrows.  (from Ref. [\cite{KVC2}])  The horizontal dashed lines show the equilibrium pressure for a given temperature, as implied by equal area construction and plotted in Fig. 4.}
\label{f1}
\end{figure}

In an implementation\cite{KVC2} of the KVC model, Anderson Hamiltonian parameters in the two phases were obtained\cite{CeTransLiu} from spectroscopic data as described further below.  To better match the experimental values of the magnetic susceptibility, the spectroscopic values of the hybridization for both phases were increased by 12\%.  Theoretical justification for such a renormalization in describing the low energy scale has been given \cite{KSAtheoryLDA3}.  After this change, there was no further parameter tuning.  The volume dependence of the hybridization is taken to be linear between the two phases, leading to a volume dependence of \TK\ and the ground state energy, evaluated using the GS method.  The non-f contribution to the bulk modulus is based on an average of the bulk moduli of La and Pr. However, the ground state energy of the Anderson model for finite \Uff\ and \nf\ $\ll$ \Nf\ is known\cite{KVC2} to include a large part that is spin independent, arising from the mixing into the ground state of the final states appearing in the BIS spectrum, f$^{1}$, f$^{2}$ and f$^3$ for La, Ce and Pr, respectively.  This spin independent contribution lowers the energy of both singlet and magnetic configurations, relative to that for zero hybridization.  There is thus an f-electron contribution to the experimental bulk moduli of La and Pr, and if this part is included in the Ce ground state energy of the Anderson model, then it must be isolated and subtracted from the bulk moduli of La and Pr so that it is not double-counted.  The volume and T-dependence of the KVC free energy can then be evaluated using the T/\TK\ dependence of the Kondo entropy\cite{Rajan} as found by the Bethe ansatz method, from which one obtains pressure/volume isotherms that imply a first order phase transition, as shown in Fig. 5.    

The resulting KVC phase boundary is shown in Fig. 4, obtained by the usual equal area construction, shown by the dashed lines in Fig. 5.  Because the input model parameters are taken from experiment and the free energy evaluation is essentially exact, the generally good agreement between the experimental and KVC phase boundaries carries the import that the KVC model correctly catches the interplay of high and low energy scales and the entropy flow.  In this calculation the entropy change is entirely magnetic with no phonon contribution, in agreement with one\cite{PhonEnt1} recent experimental result, but not another\cite{PhonEnt2}.  A feature of the phase boundary that is qualitatively important is the prediction of a second critical point at negative pressures.  This second critical point arises from the T-dependence of the Kondo entropy and was first found in the initial s-state version\cite{KVC1} of the model.   At that time the second critical point was a genuine prediction of the KVC model, and was soon verified experimentally\cite{2CeCritPoints} in alloys of Ce$_{0.9-x}$La$_x$Th$_{0.1}$, for which substitution of the smaller La ion has the effect of negative pressure, and Th substitution suppresses the magnetic $\beta$-phase. 

Fig. 6 shows 4f PES\cite{CePESData} and BIS\cite{CeBISXPSData} data for the \alf\ and \gam\ phases, along with theoretical spectra calculated\cite{CeTransLiu} using the GS method.  The dramatic change in the near \EF\ weight in the transition is very clear, although the spin-orbit sidebands of the Kondo resonance are not resolved in these wide energy range data sets.  The bulk and surface contributions to the spectra were determined by a self consistent iterative procedure\cite{CeTransLiu} that exploits the different elastic escape depths of the PES and BIS spectra, and also those of the corresponding Ce 3d PES spectra (not shown here), arising from the different photon energies used in each spectroscopy.  The energy dependence of the hybridization \DE was obtained from an LDA calculation and hence accounts approximately for the f-electrons on other lattice sites as mentioned in sub-Section 2.2.1.  The Anderson model parameters \epsf, \Uff\ and \DE\ obtained from this analysis of the spectra were used in the KVC calculation of Figs. 4 and 5, as described above.    Exactly the same parameters for bulk and surface were used to obtain the calculated spectra shown in Fig. 7, which are in excellent agreement with higher resolution data\cite{FirstCeHiRes} in which the spin-orbit sidebands are resolved.  The theory curves of the lower panels of Fig. 7 then provide a overview of the change of the entire spectrum across the transition, including the detail of the resonance and its sidebands.

\begin{figure}[!tb]
\begin{center}
\includegraphics[width=12cm]{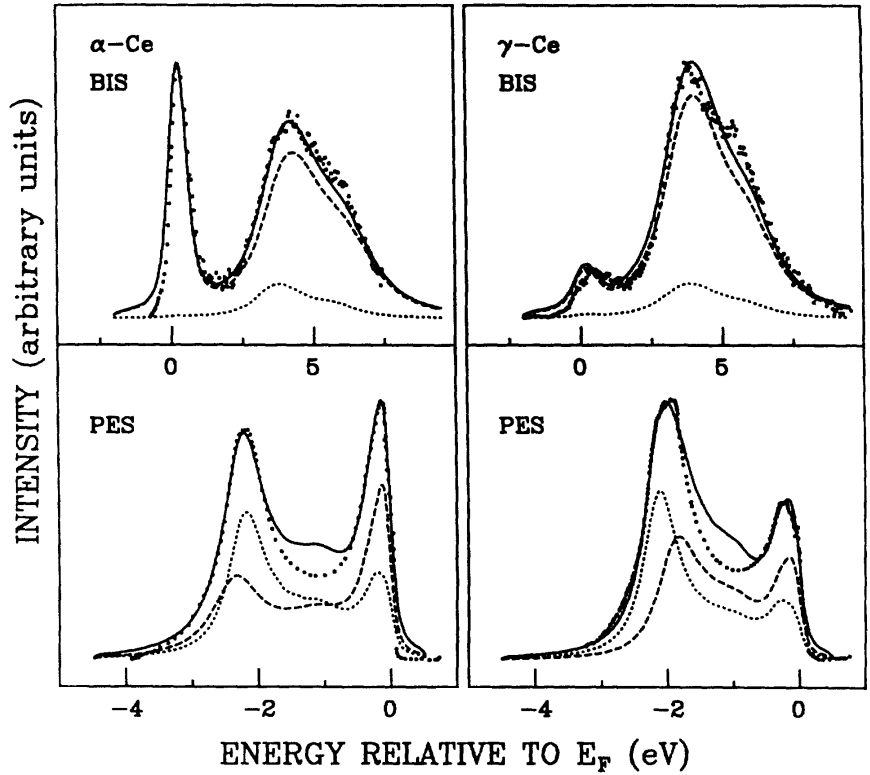}
\end{center}
\caption{Comparison of calculated (solid lines) and experimental (dotted lines) 4f PES\cite{CePESData} and BIS\cite{CeBISXPSData} spectra of \alf\ and \gam-Ce.  Surface and bulk contributions (see text) are indicated with dotted lines and dashed lines, respectively.  (from Ref. [\cite{CeTransLiu}])  The growth of the Kondo resonance from the \gam-phase to the \alf-phase is smaller but essentially like that in the data of Fig. 1 for \gam-like and \alf-like Ce materials.  Spectroscopic Hamiltonian parameters thus obtained were used for KVC calculation of Figs. 4 and 5, as described in text.}
\label{f1}
\end{figure}

\begin{figure}[!tb]
\begin{center}
\includegraphics[width=12cm]{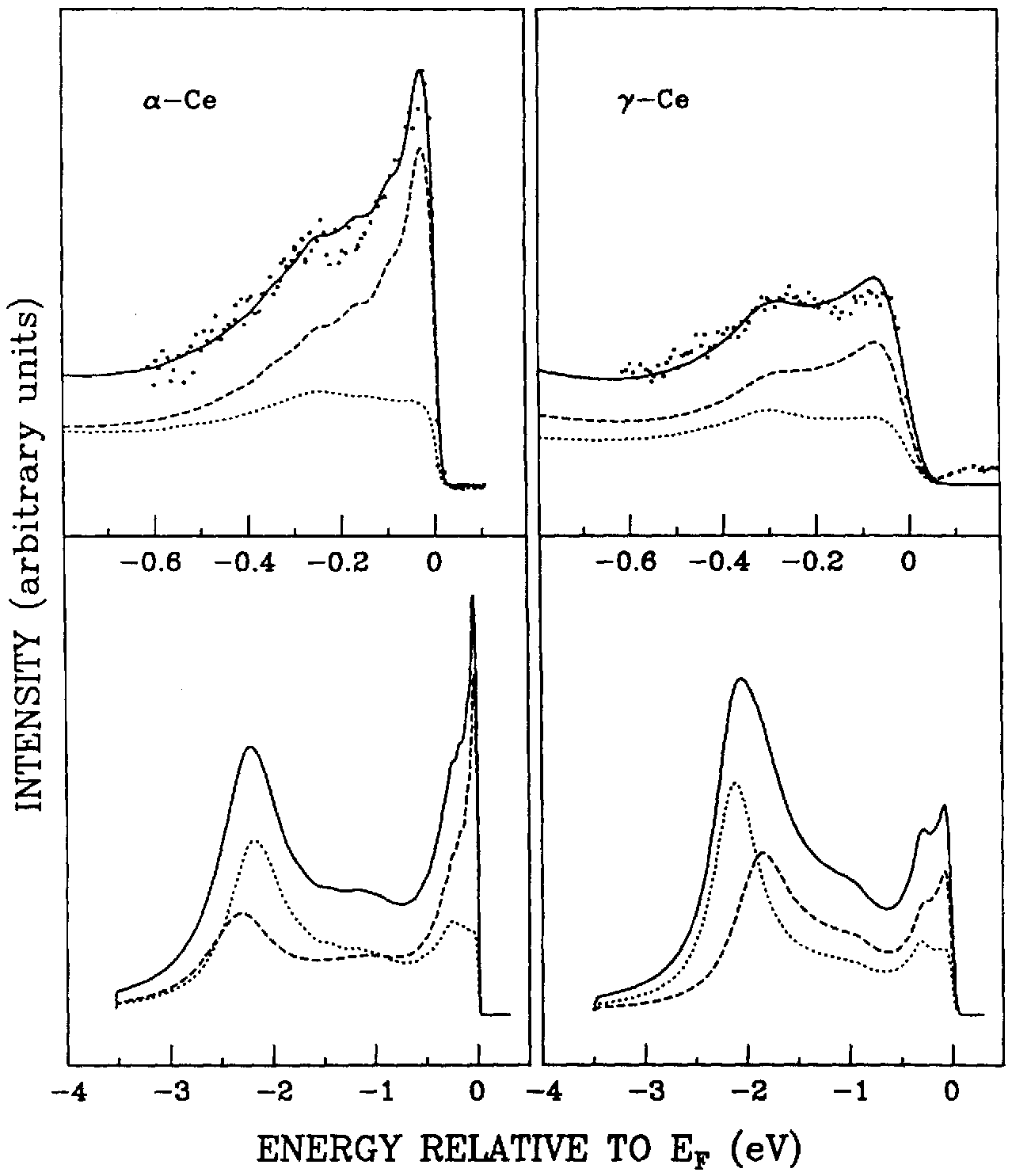}
\end{center}
\caption{Upper panels:  Experimental (dotted lines) and calculated (solid lines) 4f PES spectra\cite{FirstCeHiRes} of \alf\ and \gam-Ce with spin-orbit sidebands resolved. Calculation uses same set of Hamiltonian parameters as for Fig. 6.  Surface and bulk contributions (see text) are indicated with dotted lines and dashed lines, respectively. Lower panels:  Same theoretical spectra as in the upper panels, but plotted for the entire valence band region. (from Ref. [\cite{CeTransLiu}])} 
\label{f1}
\end{figure}

Alloying Ce with Th has essentially the same effect as applied pressure.  For the alloy Ce$_{0.7}$Th$_{0.3}$ one is effectively just beyond the critical point shown in Fig. 4, i.e. the \gam\ to \alf\ transition, as manifested by the volume and the magnetic susceptibility, is rapid but continuous over the temperature range 300K to 100K.  It is found\cite{CeThBIS} that the \EF\ resonance in the BIS spectrum also shows the transition by a rapid but smooth increase and the T-dependence of the magnetic susceptibility can be well described by the Kondo susceptibility with a smoothly varying value of \TK\ consistent with a linear volume dependence of the spectroscopic parameters, as in the KVC model.

\section{ARPES for Cerium Materials}

Early ARPES studies established the presence of \kay-dependence in the f-spectra of Ce materials\cite{Arko2, CeARPESEarly1, CeARPESEarly2, CeARPESEarly3, CeARPESEarly4}.  However it has been and continues to be a great challenge to ARPES to obtain spectra that can be compared meaningfully to theoretical calculations.  For example, essentially all knowledge of the FS has been obtained from dHvA measurements.  This section summarizes briefly some relatively recent ARPES results\cite{CeRuSiARPES} for the heavy Fermion compound CeRu$_2$Si$_2$ that show both the promise and the challenges of such studies.  The spectra shown are for cleaves believed to yield sub-surface Ce, as discussed at the close of Section 4.  Recent ARPES work on CeNi$_2$Ge$_2$ is also of note\cite{CeNi2Ge2ARPES}.

The iso-structural compounds LaRu$_2$Si$_2$, CeRu$_2$Si$_2$ and CeRu$_2$Ge$_2$ are important as literature paradigms of dHvA studies\cite{CeRuSidHvA1, CeRuSidHvA2, CeRuSidHvA3} showing that the Luttinger theorem is fulfilled for Ce f-electrons.  The essential dHvA findings are summarized in the (a) and (b) panels of Fig. 8.  The (b) panel of Fig. 8 shows a cross-section of the LDA FS\cite{Ce/LaRuSiLDA1, Ce/LaRuSiLDA2} for the La and Ce compounds, with hole (electron) pockets shown as bold (thin) lines.  The one f-electron per Ce of the Ce compound is contained about 0.5 electron each in the large hole pocket centered around the Z point and in a multiply connected electron sheet centered around $\Gamma$.  These are the high mass parts of the FS that lead to the large specific heat of this heavy Fermion material.  The (a) panel shows the large hole piece, which in the Ce compound is known as the "pillow." Its shape and reduced size in the Ce compound relative to that in the La compound is due to the added 0.5 electron/Ce.  This expected size difference has been observed in dHvA studies.  Further, CeRu$_2$Si$_2$ is metamagnetic and it is found that across the metamagnetic transition the dHvA FS switches abruptly to that of the La compound, indicating that the f-electron has been forced out of the FS volume by the appearance of the Ce magnetic moment.  Similarly, CeRu$_2$Ge$_2$ is a ferromagnet and its dHvA FS is again like that of LaRu$_2$Si$_2$.

\begin{figure}[!tb]
\begin{center}
\includegraphics[width=15cm]{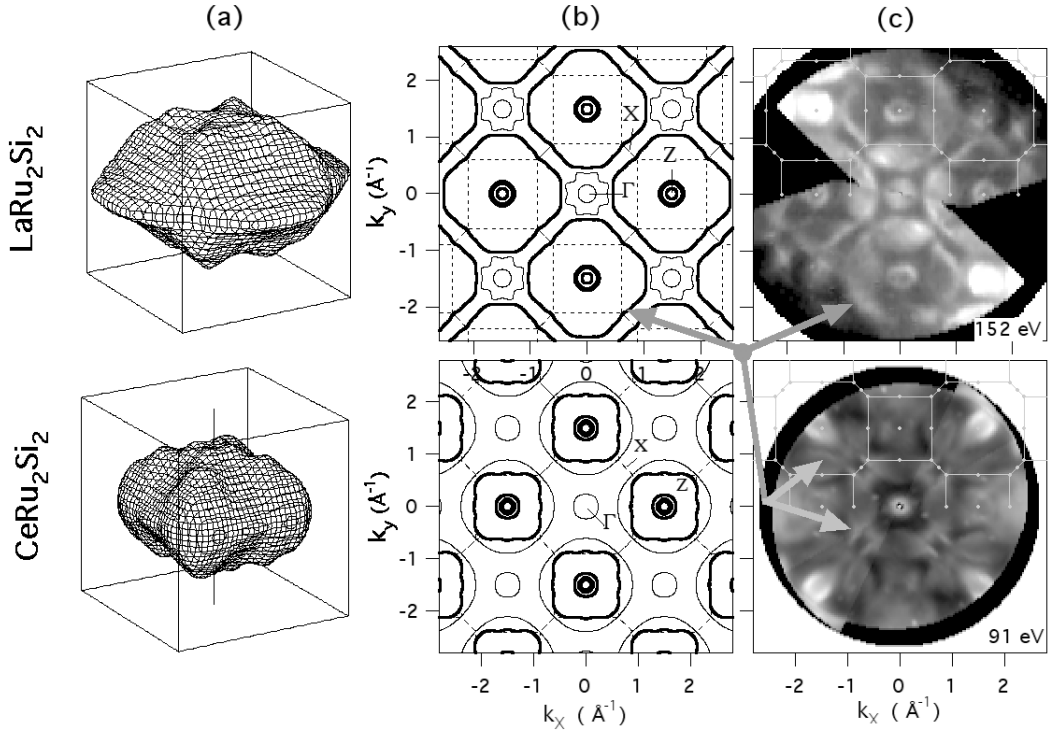}
\end{center}
\caption{ARPES Fermi surface map(lower panel (c)) for T $=$ 120K $\gg$ \TK\ $=$ 20K of CeRu$_2$Si$_2$ compared to ARPES map (upper panel (c)) of LaRu$_2$Si$_2$, showing same size of large Z-point hole surface contour (upper panels (a), (b)) for both compounds.  Absence of reduced hole surface (lower panels (a), (b)) size due to Ce 4f electron inclusion, as predicted by LDA and observed at very low T in dHvA experiments, is evidence of 4f electron exclusion from Fermi surface at T $\gg$ \TK, as described in text.  (from Ref. [\cite{CeRuSiARPES}])}
\label{f1}
\end{figure}

\begin{figure}[!tb]
\begin{center}
\includegraphics[width=15cm]{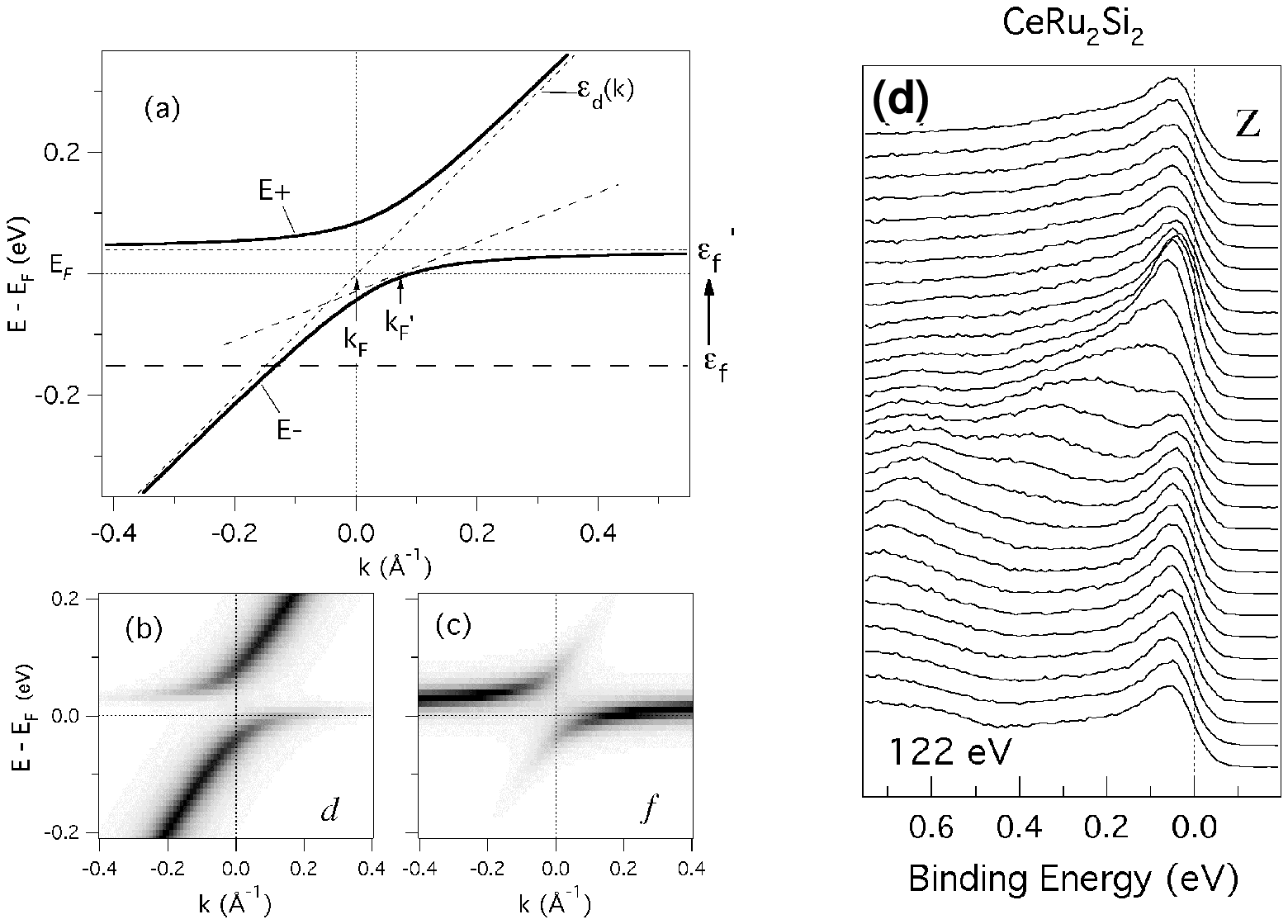}
\end{center}
\caption{Panel (a) shows schematic of the Anderson lattice renormalized two-band mixing picture of f-d hybridization near a d-band \EF\ crossing, and panels (b,c) show images of implied f and d weights along two energy branches.  Panel (d) shows high-resolution on-resonance ARPES spectra for a small low mass Fermi surface pocket near the Z-point for CeRu$_2$Si$_2$.  Fast dispersing peak is the d-band crossing \EF\ and near \EF\ weight revealed in on-resonance spectrum is of 4f character, qualitatively very much like the two-band mixing model. (from Ref. [\cite{CeRuSiARPES}])}
\label{f1}
\end{figure}

The (c) panels of Fig. 8 show ARPES FS maps for LaRu$_2$Si$_2$, and for CeRu$_2$Si$_2$ at a temperature of 120K, well above its \TK\ of 20K.  As mentioned in Section 2, it has been conjectured\cite{LuttKondoHiT1, LuttKondoHiT2} that for T well above \TK, the f-electrons should be excluded from the FS because their magnetic moments are no longer quenched.  The upper panel (c) of Fig. 8 shows the FS map obtained by ARPES for LaRu$_2$Si$_2$ at a photon energy corresponding to the \kayperp of the cross-section of panel (b).  The ARPES map is in good general agreement with the LDA and dHvA FS as to the general locations and sizes of the various FS sheets, although individual small concentric sheets around Z and $\Gamma$ are not resolved.  It is serendipitous that the large Z-point hole pocket is both of great interest and large enough that it can be resolved.  That this FS piece is indeed a hole pocket is known from the measured dispersions (not shown) of the bands that define it.  As shown in the lower panel (c) for CeRu$_2$Si$_2$, for T $\gg$ \TK\ the "pillow" that is seen in low T dHvA data is not present in the high T ARPES map.  Instead, the large Z-point FS is essentially the same in both the La and Ce compounds, providing strong evidence that the f-electrons are indeed excluded from the FS volume for T well above \TK.  The same conclusion has been reached for CeRu$_2$Si$_2$ by the less direct method of measuring the two-dimensional angular correlation of positron annihilation\cite{CeRuSiACAR}.  One would like to observe directly by ARPES the evolution of the high T FS into the low T FS, but so far ARPES measurements at low enough T have not been made.  One might speculate that the change would occur by a gradual transfer of spectral weights between the relevant places in \kay-space.

The angle integrated spectrum of CeRu$_2$Si$_2$ shows a typical Kondo resonance behavior at 120K.  It is then interesting to observe where in \kay-space the \EF\ f-electron weight is found.  The experimental answer\cite{CeRuSiARPES} from resonant ARPES is that the f-weight in the high T spectrum is in low mass regions of the FS, e.g. the small pockets at the Z-point.  At present there is no theoretical guidance as to whether this result is consistent with expectations for the Anderson lattice model, although the dispersing spectra have the generic appearance found in the simple renormalized two-band mixing picture\cite{latAnd1, latAnd2, latAndJarrell} for the Anderson lattice, mentioned in sub-Section 2.2.3.  As illustrated schematically in panel (a) of Fig. 9 this picture features an f-level renormalized from \epsf\ to a position just above \EF, rather like the Kondo resonance, and then mixing by a renormalized hybridization with a dispersing conduction band (labeled "d" in the figure). The mixing moves the \EF\ crossing \kay-vector so that the FS volume changes to contain the f-electron and makes the \EF\ mass heavier. The implied spectral distribution of f and d weight near the \EF\ crossing is shown in panels (b)and (c).  One sees that the f-weight is almost non-dispersive, lies very close to \EF\ and at non-zero T will have the largest PES amplitude inside the d-band hole pocket.  It is intriguing that the resonant ARPES spectra of CeRu$_2$Si$_2$ for a small mass hole pocket at the Z-point, shown in panel (d), greatly resemble this two-band mixing picture.  Similar spectra have been observed\cite{CeRuSiARPES} in the 5f heavy Fermion compound URu$_2$Si$_2$.  It would be very interesting to compare such ARPES data with LDA + DMFT calculations for CeRu$_2$Si$_2$, but this remains for the future.

\section{PES for \vo}

As discussed in Section 2.2.2, the transition between the PM and PI phases of \vcro\ serves as a paradigm of the Mott metal to insulator transition.  LDA + DMFT calculations  for \vcro\ predict a Kondo-like peak near \EF\ for the PM phase  In spite of careful efforts by many researchers over many years\cite{Sawatzky, Smith88, Smith94, Shin, Kim, Ralph, Schramme}, such a peak has been observed only recently\cite{MoPMphase}, through the use of high photon energy and a small diameter photon spot.  Both are important for minimizing surface effects like those described for Ce in Sections 4 and 5, i.e. that atoms on a cleaved surface, or in geometries with further reduced coordination such as steps and edges, have smaller hybridization and hence are more strongly correlated with a reduced Kondo peak.  Early PES spectra for \vo\ taken at high photon energy hinted at an important difference in the bulk and surface spectra, but the resolution was not adequate to resolve the structures in the spectrum\cite{Park}. Similar conclusions have been reached for other vanadium compounds\cite{Maiti, sekiyama2}.  Presented here are results from a high photon energy bulk sensitive PES study\cite{MoPMphase} of \vo\ performed with good resolution at beamline\cite{BL25SU} BL25SU of the SPring-8 synchrotron.

\begin{figure}[!tb]
\begin{center}
\includegraphics[width=12cm]{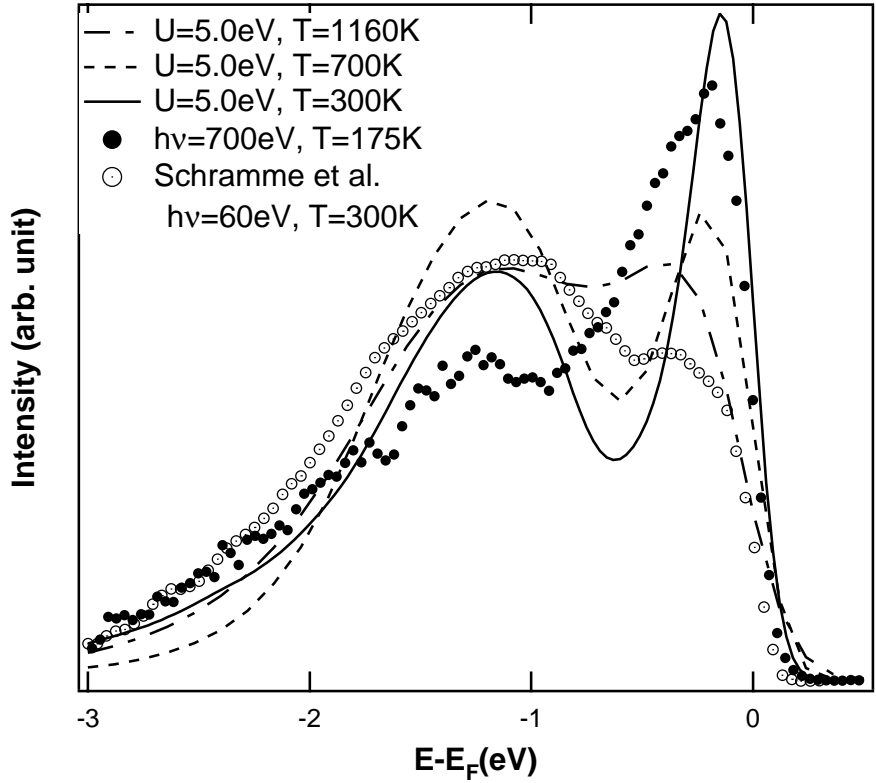}
\end{center}
\caption{LDA + DMFT calculations for the PM phase of \vo\ at three temperatures, compared to PM-phase PES spectra for \hnu\ of 60eV (60 eV spectrum from Ref. [\cite{Schramme}]) and 700eV.  The T=175K bulk sensitive 700 eV spectrum reveals the prominent Kondo-like peak predicted in 300K theory spectrum. Lack of peak in 300K 60eV surface sensitive spectrum is due to increased correlation of surface layer.   (from Ref. [\cite{MoPMphase}])}
\label{f1}
\end{figure}

Fig. 10 compares a bulk sensitive spectrum taken for \vo\ at \hnu\ $=$ 700 eV with a well resolved and high quality but surface sensitive spectrum\cite{Schramme} taken at \hnu\ $=$ 60 eV.  The large near \EF\ peak of the former is very clear and the large difference in the near \EF\ behaviors in the two spectra can be attributed to the more strongly correlated character of the surface relative to the bulk.  Fig. 10 also shows spectra calculated for \vo\ within LDA + DMFT at three temperatures.  The value U = 5 eV for the local Coulomb interaction was chosen because only values near 5eV can produce the metal to insulator transition for the electronic structures calculated for the actual atomic positions of the PM and PI phases in \vcro.  For the T=300K  theory curve the \EF\ quasi-particle peak is well separated from the lower Hubbard band, centered at about $-$1.25 eV.  The temperature of the experimental curve is close enough to that of the T=300K theory curve that a direct comparison of the two is valid.  Qualitatively the \EF\ peak and the lower Hubbard band of the theory curve are very similar, respectively, to the prominent \EF\ peak and the broad hump centered at $-$1.25 eV in the 700 eV spectrum.

In spite of the encouraging qualitative agreement between theory and experiment for the quasi-particle peak, nonetheless both the width and integrated spectral weight of the experimental \EF\ peak are significantly larger than those of the \EF\ peak in the theory curve. Within DMFT the difference indicates weaker correlations that could be described by using a reduced value of U.  However the transition to the PI phase would not then occur unless the U value were changed to be larger in the PI phase.  Such a difference is of course possible due to the less effective screening in the PI phase.  Although the spectroscopic U value is not found to change in the Ce \gam\ to \alf\ transition, the situation for \vcro\ is quite different because Ce has conduction electrons always present whereas \vcro\ loses all conduction electrons in the PI phase.  

\begin{figure}[!tb]
\begin{center}
\includegraphics[width=12cm]{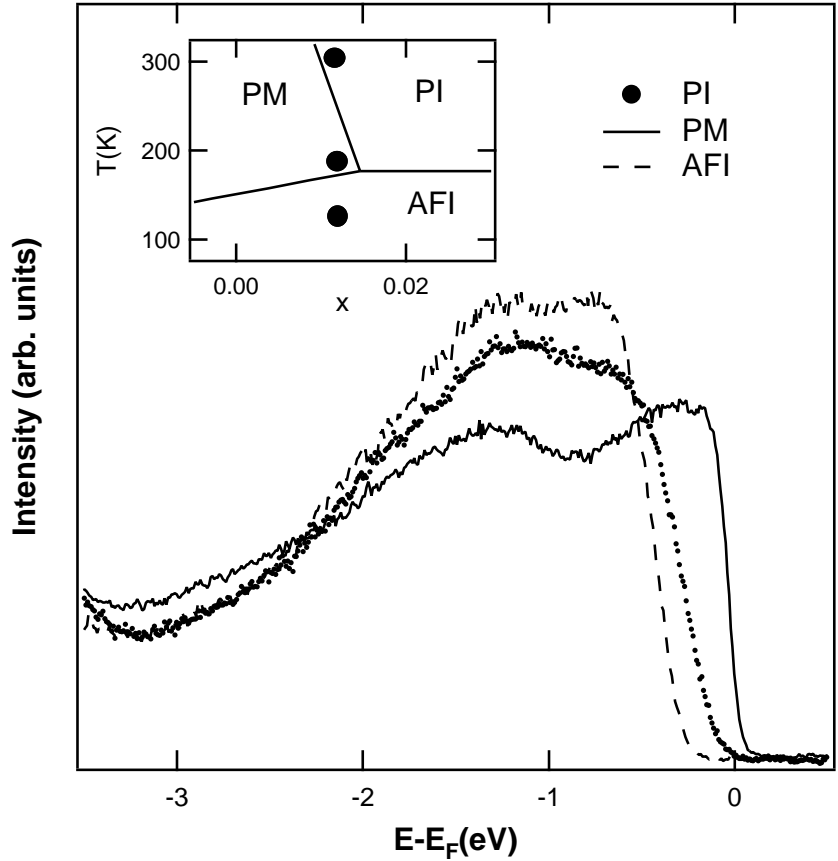}
\end{center}
\caption{Bulk sensitive PES spectra taken at \hnu\ $=$ 500 eV, showing V 3d valence band in all three phases of \vcro\ with x=1.8\%.  Gap values in PI and AFI phases are different.  Reduced weight of \EF\ peak in PM phase relative to that in \vo\ (see Fig. 10) is discussed in text.  (from Ref. [\cite{AllenIsakson}])}
\label{f1}
\end{figure}

Another possibility is that the underlying bandwidth and screening are reduced by the Cr doping in \vcro.  Fig. 11 shows the V 3d spectra\cite{AllenIsakson} obtained at \hnu\ $=$ 500 eV for a sample with x$=$1.8\%.  As shown in the inset, for this value of x all three phases can be accessed by varying T.  The \EF\ peak of the PM phase is clearly less prominent than in the \vo\ spectrum, by an amount which cannot be explained solely as a change in probe depth due to the difference in \hnu.  However the possibility of a difference in roughness for the two surfaces cannot be excluded at this time and so a clear conclusion of an intrinsic change in the PM phase peak with Cr doping cannot be made without further study to characterize the surfaces.  The opening of a gap at \EF\ in the transition to the two insulating phases is also clearly seen in the spectra.  It is interesting to notice the differing values of the gaps in the AFI and PI phases, a difference for which there is currently no theoretical understanding. 
 
The theory spectra of Fig. 10 also show that with increasing temperature the \EF\ peak loses amplitude, broadens and becomes less well separated from the lower Hubbard band.  Given the central role of the Kondo resonance in DMFT, it is not surprising that this behavior is much the same as occurs for the PES spectrum of the impurity Anderson model, as described in detail in Section 4.  As was the case for Ce it is interesting to try to observe this change, which for \vcro\ requires performing PES up to unusually high temperatures.  Thus far such a high temperature experiment has been possible only in a laboratory system, for which the low photon energy of typical high resolution laboratory sources renders the spectrum to be so surface sensitive that it does not have an actual peak at \EF.  The lack of a peak to study, and an upper temperature limit of 300K, prevented an early laboratory study\cite{Shin} of this type from drawing firm conclusions.

There is, however, another possibility which has recently been pursued with success\cite{MoPIphase}. DMFT predicts\cite{Schlipf, Bulla} that for the PI phase there will be a transfer of incoherent spectral weight from the lower Hubbard band into the gap, leading to some incoherent weight at \EF, which tends to make the material behave as a "bad metal."  This weight can be regarded as equivalent to the incoherent weight produced by the broadening of the PM phase quasi-particle peak, i.e. at high T there is no clear distinction between the incoherent correlated metal and the incoherent correlated insulator, and Cr doping x induces only a smooth cross-over behavior\cite{KotliarLandau} between the two.  Therefore another strategy for a surface sensitive laboratory spectroscopy experiment is to study the PI phase PES spectrum up to unusually high temperatures.  Since the surface is a more strongly correlated insulator than the bulk, any such gap-filling weight transfer is a valid observation. Very recently the predicted gap filling has been observed\cite{MoPIphase} and can be regarded as fitting into the general conceptual framework invoked to understand the collection of PES data on the effect of increasing temperature on the Kondo resonance in Ce materials.

\section{Conclusion}

This article has presented a combination of theory and PES and ARPES data that clearly show the beautiful Kondo resonance.  The spectroscopic data for Ce materials highlight the interplay of high and low energy scales, and the controlling role of the Kondo energy and entropy, in the Ce \alf-\gam\ transition.  PES data for \vo\ and \vcro\ serve to validate the new idea of the relevance of Kondo physics for the Mott transition.  This body of work is the product of the sustained efforts of an international community of researchers over more than twenty years time and the complete literature is even larger than was referenced in this article.  Apologies are due to the authors of many papers that are not cited.  The central role of the resonance in the impurity Anderson model, the central role of the impurity model in DMFT, and the developing broad applicability of DMFT+LDA, have combined to give the Kondo resonance a universal importance in correlated electron physics. Remaining for the future is the fascinating task of obtaining the detailed \kay-dependence and the linked T-dependence of the resonance in modern ARPES spectra for a variety of strongly correlated materials and using such data to confront the ever more sophisticated many-body theories of such materials.

\section*{Acknowledgment}

It is a pleasure to express my deeply felt gratitude to my many superb collaborators on this subject.  My research in this area is supported by the U.S. National Science Foundation under Grant No. DMR-03-02825.

\end{document}